\documentclass[prd,onecolumn,amsmath,amssymb,superscriptaddress,nofootinbib,10pt]{revtex4-2}
\usepackage[T1]{fontenc}
\usepackage{lmodern}
\usepackage{amsmath}
\usepackage{amsthm}
\usepackage{amsfonts}
\usepackage{xcolor}
\usepackage{slashed}
\usepackage{mathtools}
\usepackage{epsfig}
\usepackage[colorlinks=true,citecolor=magenta]{hyperref}

\newtheorem{conjecture}{Conjecture}

\def\({\left(}
\def\){\right)}

\def\ba{\begin{eqnarray}}
\def\ea{\end{eqnarray}}

\def\be#1\ee{\begin{eqnarray}#1\end{eqnarray}}

\newcommand{\nl}{\nonumber \\}
\newcommand{\matvec}[1]{\underline{#1}}
\newcommand{\mat}[1]{\underline{\underline{#1}}}

\begin{document}

\title{Towards a uniqueness theorem for static black holes in Kaluza-Klein theory with small circle size}

\author{Emma Albertini}
\email{emma.albertini17@imperial.ac.uk}

\affiliation{Theoretical Physics Group, Blackett Laboratory, Imperial College, London SW7 2AZ, United Kingdom}

\author{Daniel Platt}
\email{d.platt@imperial.ac.uk}

\affiliation{Theoretical Physics Group, Blackett Laboratory, Imperial College, London SW7 2AZ, United Kingdom}
\affiliation{I-X, Imperial College, London W12 0BZ, United Kingdom}

\author{Toby Wiseman}
\email{t.wiseman@imperial.ac.uk}

\affiliation{Theoretical Physics Group, Blackett Laboratory, Imperial College, London SW7 2AZ, United Kingdom}


\begin{abstract}

Kaluza-Klein theory, by which we mean vacuum gravity in 5-dimensions, with asymptotics that are a product of a circle with Minkowski spacetime, has a variety of different static black hole solutions; localized black holes and the homogeneous and inhomogeneous black strings. There is currently no uniqueness theorem for the solutions, and for fixed circle size  multiple solutions with the same mass co-exist.
Intuitively for small circle sizes we might expect the theory truncates to become 4-dimensional, and correspondingly the only black holes  are the homogeneous black strings.
Thus we conjecture that for fixed mass and sufficiently small circle size, the only black holes are homogeneous ones.
Here we give evidence that this is indeed the case.
Firstly we introduce a toy scalar field model with a potential that allows tachyonic behaviour. 
Putting this theory on a product of Minkowski with a circle gives an analogous set of static homogeneous and inhomogeneous solutions to that of the black holes. 
We prove that solutions must be homogeneous for small circle sizes, the analog of our conjecture for this toy model.
A weaker statement that is straightforward to derive is a bound on how inhomogeneous a solution can be -- putting this scalar theory in a large but finite cavity,  a norm of the wavefunction of the Kaluza-Klein modes can be shown to vanish in the small circle limit. 
Turning to the full gravitational theory, we employ a metric ansatz that imposes static axisymmetry, encompasses the homogeneous and inhomogeneous black strings (but not the localized solutions) and allows us to measure inhomogeneity of a solution. Employing a finite cavity and imposing boundary conditions that are compatible with homogeneity
we show a similar result; the norm of certain Kaluza-Klein modes is bounded by the circle size, providing evidence that our conjecture is true.

\end{abstract}

\maketitle


\section{Introduction}

Adding compact extra dimensions to spacetime has long been a route to unification, starting with the work of Kaluza and Klein, and more recently as an integral part of string theory and phenomenological scenarios such as large extra dimensions and braneworlds \cite{Arkani-Hamed:1998jmv,Antoniadis:1998ig,Randall:1999ee}. 
The exploration of gravity in dimensions greater than four revealed a far more intricate behaviour than that of conventional GR in four dimensions, even in the absence of matter fields. While linear perturbations behave in the obvious manner, black hole solutions do not, as starkly revealed by the discovery of the black ring and its associated violation of black hole uniqueness, together with even more exotic solutions~\cite{Emparan:2001wn,Elvang:2007rd,Emparan:2007wm,Emparan:2008eg}. Static black holes have a tamer nature in the asymptotically flat setting, with the natural generalization of Schwarzschild remaining the unique solution in higher dimensions~\cite{Gibbons:2002av}. However in the setting of compact extra dimensions, even vacuum static black holes have a complicated space of solutions. One may take the product of the lower dimensional Schwarzschild with a circle giving what is termed the homogeneous black string. For a given circle size, these solutions exist for any mass. However when the black hole is small it should remain unaware of the compactification and hence may be close to a higher dimensional Schwarzschild solution. Such solutions are termed localized black holes and already reveal a lack of uniqueness for small masses relative to the circle size. 

The situation became more complicated with the Gregory-Laflamme instability~\cite{Gregory:1993vy,Gregory:1994bj}. The instability has fascinating dynamical ramifications~\cite{Choptuik:2003qd,Figueras:2022zkg}, but also led to the discovery of  static inhomogeneous black strings~\cite{Horowitz:2001cz,Gubser:2001ac,Wiseman:2002zc}, and revealed a complicated space of solutions where the localized and inhomogeneous branches connect~\cite{Kudoh:2004hs, Headrick:2009pv}. These solutions have no analytic construction and exploration has relied on numerical methods  starting in~\cite{Wiseman:2002zc,Sorkin:2003ka,Kudoh:2003ki,Headrick:2009pv} and reviewed in~\cite{Wiseman:2011by,Dias:2015nua}. More recently large $D$ techniques have allowed analytic insights too~\cite{Emparan:2013moa,Bhattacharyya:2015dva,Emparan:2015hwa}. 
There are a number of reviews of this fascinating story~\cite{Kol:2004ww, Harmark:2005pp, Horowitz:2011cq}.
The violation of uniqueness is very interesting, and is related to scaling behaviour near the topology change where the inhomogeneous black strings merge with the localized black hole branch, as brilliantly predicted in~\cite{Kol:2002xz}, and revealed in exquisite numerical work~\cite{Kalisch:2015via,Kalisch:2017bin,Cardona:2018shd}. For masses close to that where the conical transition between the branches occurs, we expect to have many solutions with that same mass due to the complex exponents governing the critical merger point, and an arbitrarily large number is possible if the mass is tuned close to the critical mass, even considering only one horizon topology. Solutions are naturally characterized by both their mass and also their tension~\cite{Townsend:2001rg,Harmark:2004ch,Kol:2003if}. However unlike mass, tension is not a conserved quantity, and so while tension may be useful to characterize different solutions with the same mass, it does not give a notion of uniqueness similar to our usual understanding in 4 dimensions. 
Modifications to the notion of uniqueness applying also a criterion of stability have been proposed~\cite{Kol:2002dr,Emparan:2008eg}, but their status is unclear.

In this work we explore the idea that uniqueness in the conventional sense may exist for sufficient masses compared to the circle size. 
We focus on static black holes in vacuum Kaluza-Klein theory, by which we mean Einstein gravity in 5-dimensions where the asymptotic geometry is a product of 4-dimensional Minkowski spacetime and a circle of radius $L$. Given a black hole of mass $M$ we may define a length $R_S = 2 M$ using the usual 4d Schwarzschild radius (in natural units). 
Naively we would imagine that when the circle radius $L$ is taken to be parametrically small compared to this natural length scale of the black hole, $R_S$, then the situation should be simpler. We should recover 4-dimensional physics which is given by the homogeneous solutions, and thus for static black holes we should uniquely find the homogeneous black string. This fits with what is known of the localized black holes and black strings, namely that at high masses relative to the circle size only the homogeneous solution is known to exist. However there has been no previous work directed at showing the theory necessarily truncates to the homogeneous sector. It is the purpose of this work to initiate a study of this. 
We note that in vacuum this classical theory enjoys a scale symmetry and thus it is the invariant given by the ratio of mass to circle radius that is the relevant quantity. 
Thus we conjecture the following;
\\

\begin{conjecture}
\label{conj}
Consider smooth static black hole solutions of vacuum Kaluza-Klein theory with mass $M$ and circle radius $L$. Define the ratio $\mu = M/( 2\pi L)$, which gives the average mass per unit length. Then there exists some $\mu_0$ such that the only solutions with $\mu > \mu_0$ are homogeneous black strings. \\
\end{conjecture}

The remainder of this paper will attempt to give evidence supporting this conjecture. Firstly in section~\ref{sec:scalar} we develop a scalar field toy model living in 5 dimensions with a compact circle whose static solutions have similar behaviour to those of the Kaluza-Klein black holes.
There exist both solutions that are homogeneous on the circle, and ones that are inhomogeneous. 
In order to achieve this the theory has a potential that may be negative.
In this simple setting we give an elementary proof that for sufficiently small circle sizes (now relative to the negative energy scale in the potential) only homogeneous solutions exist -- this is the analog of our conjecture above. In order to make this proof we consider the system cut off in the radial direction, so in a `box' whose spatial boundary is a 2-sphere product with the circle. 
A weaker statement that follows almost immediately from the equation of motion for this system is that a 4-dimensional integral norm of each Kaluza-Klein mode is bounded by the circle radius. Here by Kaluza-Klein mode we mean that we decompose the scalar field in harmonics on the compact circle, and the Kaluza-Klein mode is the (static) 4-dimensional wavefunction associated to a given non-constant harmonic. As the circle size decreases this integral norm becomes smaller, and hence the Kaluza-Klein modes become smaller. It does not imply the stronger uniqueness result, but does suggest it.   

In section~\ref{sec:KalzuaKlein} we turn to showing a similar result in vacuum Kaluza-Klein theory, focussing on static black hole solutions that are spherically symmetric in the 4-dimensional sense, and hence axisymmetric in the full 5-dimensions.
Unfortunately we are unable to prove our conjecture above. However we are able to prove a version of the weaker statement we make for the toy scalar model, namely we prove that for the component of the metric that controls the spatial 2-sphere associated to the spherical symmetry, the integral norm of its Kaluza-Klein modes is bounded by the size of the circle. 
In order to make this statement we must also place the gravitational system in a  cavity. 
Hence as the circle size goes to zero, keeping certain cavity data fixed (that for a homogeneous black string fixes its mass per unit length) this norm also is forced to vanish, showing that the inhomogeneity of this metric component is vanishing, as measured in  this integral norm.
We work in Euclidean signature, where the cavity wall is then given by a product of the Euclidean time circle, a spatial round 2-sphere, and the compact spatial circle. We are also forced to consider only solutions with string topology, and thus we make no such statement about localized black holes. 
A subtle point that must be addressed is what is meant by homogeneity -- a solution that appears homogeneous having no dependence on some coordinate can be made to appear inhomogeneous by a coordinate transformation. Thus we reference the homogeneity relative to the lapse function, and hence work in a chart adapted to this. In particular we rely heavily on an analog of the remarkable chart first discussed by Harmark and Obers~\cite{Harmark:2002tr}, although not shown to be consistent until later in~\cite{Wiseman:2002ti}. Here we extend the discussion of the existence of this chart, and are able to prove that if there exists any chart that globally covers a static axisymmetric black string solution, then our adapted chart also exists.
We conclude the paper with a discussion.

\section{A scalar field  toy model}
\label{sec:scalar}

Firstly let us consider a real scalar field $\Phi$ in $d$-dimensional Minkowski spacetime with a potential $V(\Phi)$. Further we restrict to static and spherically symmetric field configurations so $\Phi = \Phi(r)$, with $r = \sqrt{ x^i x^i}$ being the usual spatial radial coordinate. Then we have the scalar field equation of motion,
\be
\label{eq:scalarsph}
\partial_r^2 \Phi + \frac{(d-2)}{r} \partial_r \Phi  - V'(\Phi) = 0 \; .
\ee
Here we will require the potential to obey the following conditions; there exist constants $C$, $C'$ such that,
\be
\label{eq:potlbounds}
\Phi V'(\Phi) \ge - C^2 \; , \quad V''(\Phi) \ge - {C'}^2 \quad \forall \; \Phi \; .
\ee
This allows a tachyonic mass, provided the mass-squared isn't too negative.

\begin{figure}[h]
\includegraphics[width=8cm]{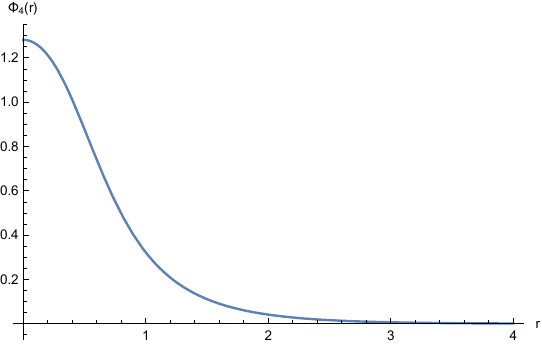} \hspace{0.5cm} \includegraphics[width=8cm]{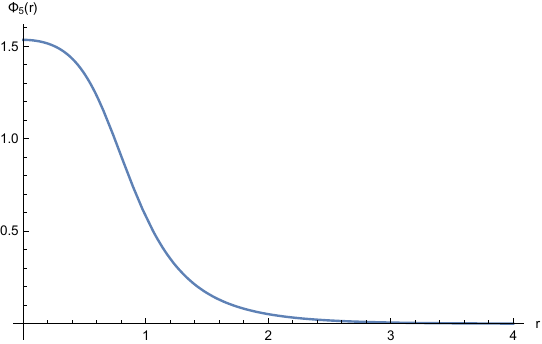}
\caption{
\label{fig:scalarprofile}
Plots showing the profile  of the scalar field for $d = 4$, the solution $\Phi_4(r)$ in the lefthand frame, and the $d=5$ solution $\Phi_5(r)$ on the right, both taking the asymptotic boundary condition that $\Phi$ vanishes. 
}
\end{figure}

We will now choose our potential $V(\Phi)$ to have the following form,
\be
\label{eq:potl}
V(\Phi) =  \Phi^2 - 4 \Phi^4 + \Phi^6 \; .
\ee
A static solution is given by $\Phi = 0$, and the theory has massive stable excitations about this perturbatively. Non-perturbative deformations of the scalar from zero may probe the negative region of the potential. This potential obeys the conditions above, where one can compute that the best $C$ and $C'$ are,
\be
C^2 = \frac{4 \left( 404 + 55\sqrt{55} \right)}{243}  \; , \quad C'^2 =  \frac{86}{5}  \; .
\ee
In both $d=4$ and $d=5$ dimensions we may find localized static spherically symmetric solutions where the scalar asymptotes to zero for large radius. This may be regarded as a boundary value problem; we require regularity at the origin (so the scalar is a smooth function of $r^2$) and asymptotic decay. While there are multiple solutions, we focus on the solutions which have no nodes. 
We denote these solutions $\Phi_4(r)$ and $\Phi_5(r)$ for the cases $d = 4$ and $5$ respectively. They may be straightforwardly found numerically and are shown in figure~\ref{fig:scalarprofile}. 
We expect all such non-zero scalar solutions to be unstable as their existence is not topologically protected.
We further expect the solutions with more nodes to be increasingly unstable. We emphasize that while the nodeless solutions we focus on should be unstable, they are nonetheless valid static solutions, and will give us a useful toy model of the Kaluza-Klein black holes as we now describe.

Let us now take the same theory, but on the Kaluza-Klein vacuum spacetime, so the product of Minkowski with a circle of radius $L$. We will use coordinates $(x^\mu, \theta)$ where $x^\mu$ are the usual 4-dimensional Minkowski coordinates, $x^\mu = (t, x^i)$ and $\theta$ is an angle which covers the compact fifth  dimension, so $\theta \sim \theta + 2 \pi$.
In order to simplify the discussion of asymptotic boundary conditions, instead we will take the spacetime to be the interior of a cavity with 4-dimensional radius $R$, so in our coordinates the cavity wall is located at $| x^i |  = R$ and is the product of time with the round sphere of radius $R$ and the compact circle.
We  place boundary conditions for the field on the cavity wall.
We then write the metric as,
\be
ds^2 = \eta_{\mu\nu} dx^\mu dx^\nu + L^2 d\theta^2 
\ee
for $| x^i |  \le R$. Now we take $\Phi$ to be a real static scalar field on this background, so that, $\Phi = \Phi(x^i, \theta)$ and it is periodic in $\theta$. Its equation of motion is,
\be
\label{eq:scalarEOM}
\partial_i \partial_i \Phi + \frac{1}{L^2} \partial_\theta^2 \Phi - V'(\Phi) = 0
\ee
for the same potential as above in equation~\eqref{eq:potl}. We take Dirichlet boundary conditions at the cavity wall, so that the field vanishes there, $\Phi|_{r = R} = 0$. 
We may decompose a solution into a homogeneous component $\bar{\Phi}(x^i)$ -- the `zero mode' --  and an inhomogeneous one, which we write in terms of  the non-constant Fourier modes -- the `Kaluza-Klein modes' -- on the circle $\Phi_n(x^i)$ for $n \ne 0$; hence,
\be
\Phi(x^i, \theta) = \bar{\Phi}(x^i) + \sum_{n \ne 0} \Phi_n(x^i) e^{i n \theta}
\ee
with the reality condition $ \Phi^\star_n(x^i) =  \Phi_{-n}(x^i)$. Each Kaluza-Klein mode must separately vanish at the cavity wall, as must the zero mode.

Let us assume that the cavity radius is reasonably large compared to the scales in the potential which are $\sim O(1)$, so $R \gg 1$. In that case, for any circle size $L$, we should find a solution close to the solution $\Phi_4(r)$;
\be
\Phi(x^i, \theta) \simeq \Phi_4(r ) \; .
\ee
This is approximate since we have changed the boundary condition from $\Phi \to 0$ as $r \to \infty$ now to $\Phi = 0$ at $r = R$. However provided $R$ is quite large the change to the solution is very small. 
More precisely, assuming our solution is homogeneous on the circle, so $\Phi$ is independent of $\theta$, then the equation of motion for a static spherically symmetric solution $\Phi = \Phi_{homog}(r)$, simply reduces to~\eqref{eq:scalarsph} with $d=4$, with the boundary condition that $\Phi$ vanishes at $r = R$.
We will illustrate our discussion by choosing the example cavity radius of $R = 4$. While this value of $R$ is not very large, if we were to superpose $\Phi_{homog}(r)$ on the left hand frame of figure~\ref{fig:scalarprofile}, then by eye we would see no difference to the case $\Phi_4(r)$ which decays asymptotically and is shown there.
Thus we have a class of solutions which are independent of $\theta$, so homogeneous on the circle, and as a result exist for any $L$. To form the analogy with the Kaluza-Klein black holes, these correspond to the homogeneous black strings. We should think of this solution $\Phi$ as the analog of a black string of fixed mass per unit length along the circle; this black string solution then exists for any circle size, and hence mass.

The localized black holes have an analog solution too.  In the limit that both $R$ and $L$ are large, $1 \ll R, L$, then a localized solution centred at $x^i = 0$, $\theta = \pi$ is approximately given by the 5-dimensional spherically symmetric solution $\Phi_5$ above as,
\be
\Phi(x^i, \theta) \simeq \Phi_5(\rho) \; , \quad  \; \rho = \sqrt{ x^i x^i + L^2 (\theta-\pi)^2 }  .
\ee
Now for our reasonably large $R = 4$, and for large $L$, we expect the solution to still exist, albeit slightly deformed from its asymptotically decaying form. This is the analog of a localized black hole in a very large extra dimension, so that to a good approximation it appears to be a higher dimensional Schwarzschild solution. Just as in that case, as the size of the circle $L$ is reduced, we expect a solution for $\Phi$ will persist but will become deformed as it `sees' itself around the circle.
As for the black strings and black holes both the homogeneous and inhomogeneous solutions described above are axisymmetric solutions. Hence they can be written as $\Phi = \Phi(r, \theta)$ which solves,
\be
\label{eq:scalarEOMsym}
\partial_r^2  \Phi + \frac{2}{r} \partial_r \Phi + \frac{1}{L^2} \partial_\theta^2 \Phi - V'(\Phi) = 0
\ee
with $\Phi$ being a smooth function of $r^2$ at the axis of symmetry $r = 0$.

Interestingly the same phenomenon occurs for these localized solutions as for the localized black holes. As $L$ is reduced, and the solutions become larger relative to the circle size, they become increasingly homogeneous until at a critical value, $L_{crit}$, they become exactly homogeneous. This implies that for $L = L_{crit}$ the homogeneous solution must have a static deformation, the analog of the Gregory-Laflamme marginal mode. Indeed just as for the black string, precisely for $L > L_{crit}$ this marginal mode flows to become a dynamical instability of the homogeneous solutions -- the analog of the Gregory-Laflamme instability. 
We may solve the  full time dependent scalar equation of motion by perturbing about the homogeneous solution $\Phi_{homog}(r)$ as,
\be
\Phi(t, x^i, \theta)  = \Phi_{homog}(r) + \epsilon\, \delta \Phi_n(r) \cos( n \theta ) e^{i \omega t} + O(\epsilon^2)
\ee 
so that linearizing in $\epsilon$, we find,
\be
\partial_r^2 \delta \Phi_n + \frac{2}{r} \partial_r \delta \Phi_n  + \left( \lambda + 48 \Phi_{homog}(r)^2 - 30 \Phi_{homog}(r)^4 \right) \delta \Phi_n = 0 \;  , \quad \lambda =  \omega^2 - 2 - \frac{n^2}{L^2} \; .
\ee
As for the Gregory-Laflamme instability, this must be solved for regularity of $\delta \Phi_n$ at the origin, and vanishing at the cavity wall $r = R$, and then it determines the allowed values of $\omega^2$ for a given circle size $L$ and harmonic $n$. However for this toy case the dependence on $L$ and $\omega$ is much simpler being determined by the quantity $\lambda$ above, whose value may be found numerically to be $\lambda = -8.19$.
Thus we see that since $\omega = 0$ and $n = 1$ for $L = L_{crit}$ for the marginal mode, then we must have, $L_{crit} = 0.402$. The time dependent modes are then given by,
\be
\omega^2 = \frac{n^2}{L^2} - \frac{1}{L_{crit}^2} \; .
\ee
Thus this instability exists for $L > L_{crit}$, just as for the Gregory-Laflamme instability. However, unlike the Gregory-Laflamme case, $\omega^2 \to - 1/L_{crit}^2$ as $L$ becomes very large, rather than going to zero, reflecting the fact that the 4-dimensional solution $\Phi_4(r)$ is unstable.

\begin{figure}[h]
\includegraphics[width=5.5cm]{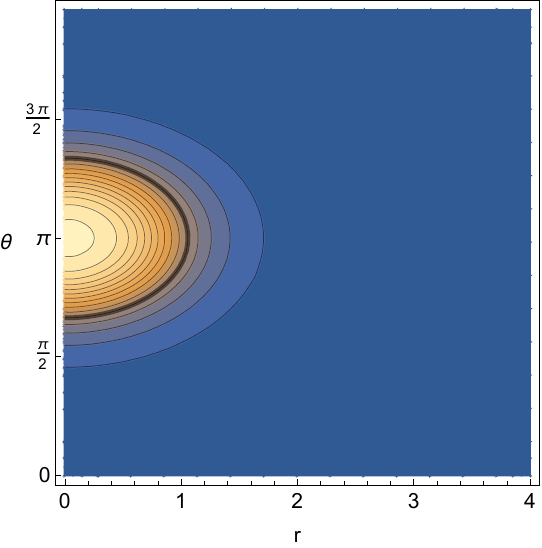} \hspace{0.3cm} \includegraphics[width=5.5cm]{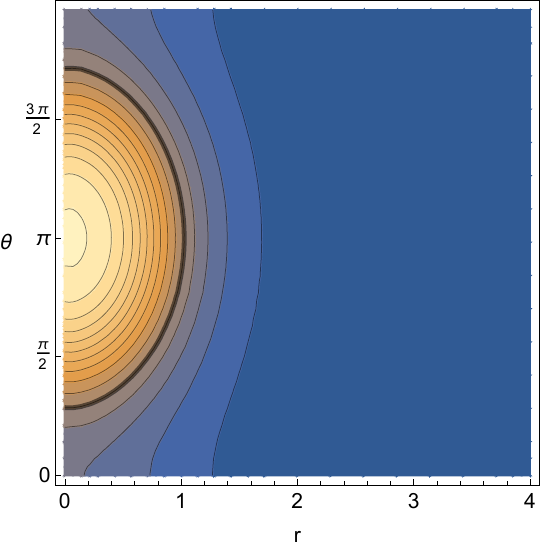} \hspace{0.3cm} \includegraphics[width=5.5cm]{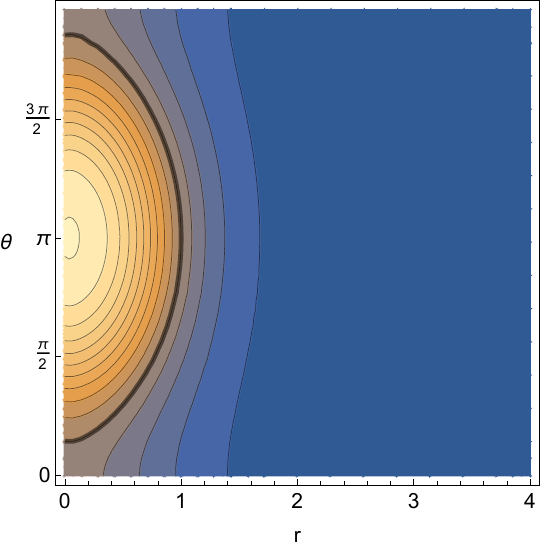} \\
\vspace{0.5cm}
\includegraphics[width=5.5cm]{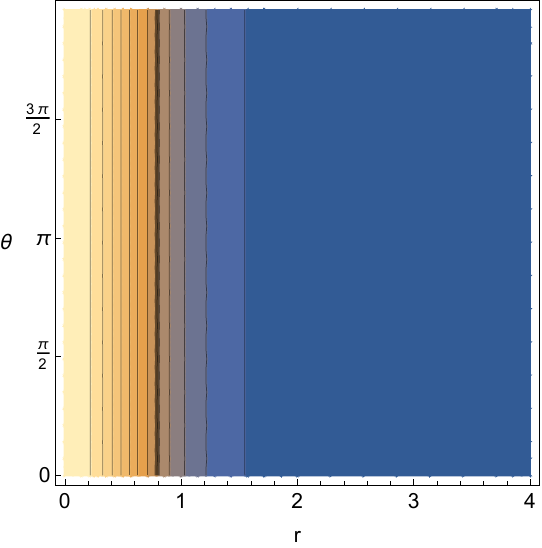} \hspace{0.3cm} \includegraphics[width=5.5cm]{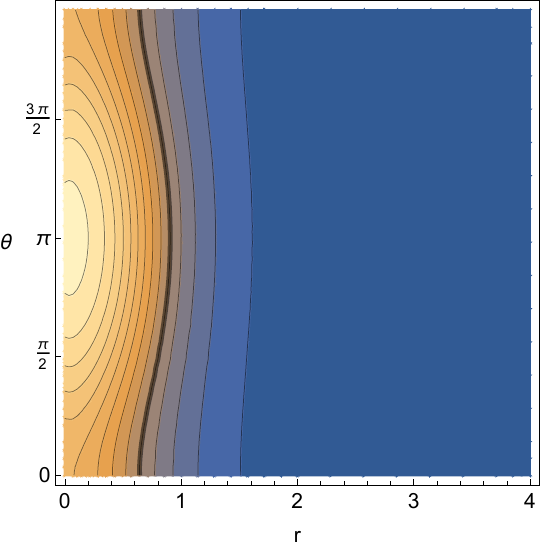} \hspace{0.3cm} \includegraphics[width=5.5cm]{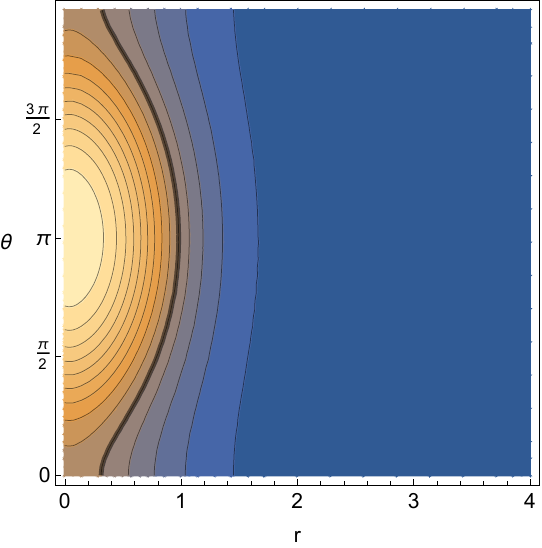}
\caption{
\label{fig:scalarsolutions}
Plots showing static axisymmetric scalar field solutions for various values of the circle size. The field is set to zero at the cavity wall located at $r = 4$. The top left plot is for $L = 1$ and we see the solution is localized on the circle, and appears similar to the asymptotically flat 5-dimensional solution $\Phi_5$. Moving clockwise the circle size decreases through the values $\{ 0.5, 0.46, 0.44, 0.41, 0.4 \}$, and we see the solution become increasingly homogeneous, until the critical value $L_{crit} = 0.402$ when the solution becomes exactly homogeneous. The contours plotted are in intervals of a tenth, with the contour whose value is a half being shown as a heavier line. This contour transitions from being spherical in the 4 spatial dimensions to becoming cylindrical, wrapping over the circle. We see the static solutions of this scalar system have an analogous behaviour to that of the black strings in Kaluza-Klein theory.
}
\end{figure}

Using elementary numerical methods we may explicitly find these axisymmetric scalar field solutions governed by~\eqref{eq:scalarEOMsym}, and we show a selection of these for various $L$ in figure~\ref{fig:scalarsolutions}. These demonstrate the transition from a localized solution with relatively large $L = 1$, whose form is approximately spherically symmetric, through increasingly homogeneous solutions, until the solutions become exactly homogeneous for $L < L_{crit}$.
In order to distinguish between an analog of the  localized black hole and that for the inhomogeneous black strings, we may consider a level set of the scalar. As we see from the figure, if we draw the level set, say at $\Phi = 1/2$ then this does have a transition between a spherical topology for large $L$ -- analog to the localized black hole -- to a cylindrical topology for smaller $L$ -- the analog of the inhomogeneous black string. 
With this toy model in hand, we now investigate what statements we may make about the existence of inhomogeneous solutions when the circle radius $L$ becomes small.

\subsection{Integral bound on Kaluza-Klein modes} 

If we take the static scalar equation of motion~\eqref{eq:scalarEOM} and integrate it over space against the spatial measure and the field, then after integration by parts we obtain;
\be
0 = L  \int dx^i d\theta  \left( ( \partial_i \Phi)^2 + \frac{1}{L^2} (\partial_\theta \Phi)^2 + \Phi V'(\Phi) \right) - \int_{r=R} dy^a d\theta \sqrt{h} \Phi \partial_n \Phi
\ee
where the metric induced on the cavity surface is,
\be
ds_{r = R} = h_{ab} dy^a dy^b = -dt^2 + R^2 d \Omega_{(2)}^2 + L^2 d\theta^2 
\ee
and $\partial_n$ is the unit outward directed normal derivative to the cavity, and $d \Omega_{(2)}^2$ is the line element on the round unit 2-sphere.
Now since our boundary condition is that $\Phi = 0$ on the cavity wall, the surface term vanishes, and we obtain,
\be
  \int dx^i d\theta    (\partial_\theta \Phi)^2   &=& -  \int dx^i d\theta \left( L^2 \Phi V'(\Phi) + L^2 ( \partial_i \Phi)^2 \right) \le L^2 C^2 \int dx^i d\theta = \left(  \frac{8}{3} \pi^2 C^2 R^3 \right) L^2 
\ee
 using the bounds on the potential in~\eqref{eq:potlbounds}. Substituting in the Kaluza-Klein decomposition we obtain,
\be
4 \pi \sum_{n > 0} n^2 \int dx^i  \left| \Phi_n(x^i) \right|^2   \le \left(  \frac{8}{3} \pi^2 C^2 R^3 \right) L^2 
\ee
so that a consequence of this is simply that each Kaluza-Klein mode must individually obey an upper bound on its $\mathcal{L}_2$ norm,
\be
\label{eq:KKbound}
\left( \mathcal{L}_2\left[ \Phi_n \right] \right)^2 \equiv \int dx^i  \left| \Phi_n(x^i) \right|^2   \le \left(  \frac{2 \pi C^2 R^3 }{3 n^2} \right) L^2  \; .
\ee
Now we regard the potential, and hence $C$, as fixed, as well as the cavity radius $R$. Then if we take $L$ to be small we see that this norm of $\Phi_n(x^i)$ becomes small, going linearly with $L$ to zero.
Hence we see that as the circle size becomes small, the solution must become increasingly homogeneous, at least in the sense of this  $\mathcal{L}_2$ norm for the Kaluza-Klein modes.

\subsection{Proof of no inhomogeneous solutions}

For this simple example of the scalar field we may go further and prove that there exist no inhomogeneous solutions for circle sizes less than some minimum value $L_{min}$. Firstly we take a derivative of equation~\eqref{eq:scalarEOM} with respect to the circle angle, and then defining $\psi = \partial_\theta \Phi$ this yields,
\be
\partial_i \partial_i \psi + \frac{1}{L^2} \partial_\theta^2 \psi - V''(\Phi) \psi = 0 \; .
\ee
 Now integrating this over space against the spatial measure and $\psi$, and integrating by parts then we obtain,
\be
0 & = &  L  \int dx^i d\theta  \left( ( \partial_i \psi)^2 + \frac{1}{L^2} (\partial_\theta \psi)^2 +  \psi^2 V''(\Phi)  \right) -  \int_{r=R} dy^a d\theta \sqrt{h} \psi \partial_n \psi
\ee
and we note that again the surface term vanishes as $\Phi$ is zero, so constant on the cavity wall; hence $\psi = 0$ there. Thus using our bounds on the potential~\eqref{eq:potlbounds} we learn that,
\be
\int dx^i d\theta  \left(  \frac{1}{L^2} (\partial_\theta \psi)^2 -  \psi^2 {C'}^2  \right) \le 0 
\ee
and inserting the expansion in Kaluza-Klein modes we find,
\be
\sum_{n > 0} n^2 \int dx^i   \left(  \frac{n^2}{L^2} - {C'}^2 \right)  \left| \Phi_n(x^i) \right|^2  \le 0 \; .
\ee
This implies that if the radius $L$ is too small, the only solution of the inequality above is that all the non-trivial Kaluza-Klein modes $\Phi_n(x^i)$ must vanish and hence the solution is homogeneous;
\be
\frac{1}{L^2} \ge  {C'}^2 \quad \implies \quad \Phi_n(x^i) = 0 \; .
\ee
We see that it is necessary that $L > 1/C'$ for a solution to be inhomogeneous; this then shows that, $L_{min} > 1/C'$. For our example potential  this gives $L_{min} \gtrsim 0.241$. We have seen above that the inhomogeneous solutions constructed earlier exist only for $L > L_{crit} = 0.402$ which is indeed consistent with this bound.
 
The Kaluza-Klein modes behave as 4-dimensional massive fields and hence decay with an exponential Yukawa fall-off. It is then worth noting that since $\partial_\theta \Phi$ decays exponentially at large radius, the proof above still stands even if we take the cavity size to infinity. However our simple bound on the norm of the Kaluza-Klein modes does not survive this limit; if we keep $L$ fixed as $R \to \infty$ then the righthand side of the bound diverges.
 
\section{Kaluza-Klein theory}
\label{sec:KalzuaKlein}

We now turn to the Kaluza-Klein theory, namely vacuum 5-dimensional gravity with asymptotics that are a product of Minkowski spacetime with a circle of radius $L$. We will not be able to consider the theory for general static solutions. Here instead we will restrict to solutions that are static and axisymmetric, being spherically symmetric in the extended spatial dimensions, and that have a string topology horizon. Further we will require  solutions to have a mirror reflection symmetry about the circle. We emphasize that all the known static black hole solutions in Kaluza-Klein theory fall in this class.
Again, in order to simplify the discussion of asymptotics we will place the system in a cavity, and we will impose homogeneous boundary conditions on the cavity wall. Provided  that the radius of this cavity is quite large compared to the horizon radius of solutions, we expect that the usual asymptotically Kalzua-Klein black hole solutions again exist, albeit slightly deformed by the existence of the cavity wall. However we expect a richer space of static solutions, with new solutions that cannot exist with the usual asymptotics. For example, the homogeneous black string that is asymptotically Kalzua-Klein is a 4-d Schwarzschild solution product with a circle, and hence in a big cavity will have a new `large' black string counterpart in the sense of York, Hawking and Page~\cite{York:1986it,Hawking:1982dh}. There may also be counterparts which are inhomogeneous too. The important point for us is that provided the cavity is reasonably large compared to the black hole radius, all our usual solutions of interest will be captured. The question is then what happens to inhomogeneous solutions as the circle size is made very small, as compared to the black hole size (and also the cavity radius)?

\subsection{Black string ansatz}

Since we will be interested in static solutions it is convenient to treat the system by the usual continuation to Euclidean signature so that the metric takes the form,
\be
\label{eq:static}
ds^2 = r(x)^2 d\tau^2 + g_{ij} dx^i dx^j 
\ee
where $r(x^i)$ is the lapse function. The coordinate $\tau = i t$ is Euclidean time. 
Since we are interested in black hole solutions we take this coordinate to be periodic with period $\beta$, so that $\tau \sim \tau + \beta$. Then as usual, in asymptotically flat spacetime $\beta$ is a measure of the inverse temperature of the solutions; in the setting of a cavity it provides a measure of the temperature of the cavity. The Lorentzian static symmetry implies that this Euclidean metric has a $U(1)$ isometry generated by the Killing vector $\partial/\partial \tau$.  The horizon -- by which we mean the Euclidean analog of the horizon, the fixed point set of the $U(1)$ isometry --  is where $r(x) = 0$.
We will define the cavity wall by requiring that $g_{\tau\tau}$ is constant there. By rescaling the $\tau$ coordinate along with its period $\beta$, we may adjust this constant to take any value. Hence we will choose its value to be one, so that $r(x) = 1$ at the cavity boundary. Then the Euclidean time circle has length $\beta$ at the boundary, and the condition that $g_{\tau\tau} = 1$ will be our first boundary condition that we impose. We note that we may view this as a constant gravitational potential condition at the cavity boundary, and emphasize that it is compatible with having homogeneous solutions in the interior.

We then make a number of key assumptions. Firstly we assume that the Euclidean geometry is of black string topology and is smooth with a regular (Euclidean) horizon. Secondly, we assume that this static metric in the exterior of the horizon is axisymmetric, being spherically symmetric in the extended spatial directions (which are now truncated by the cavity). Thirdly, we assume that there exists \emph{some} coordinate chart that globally covers the solution as,
\be
\label{eq:generalmetric}
ds^2 = r(\rho,z)^2 d\tau^2 + e^{2 U(\rho,z)} d\rho^2 + e^{2 W(\rho,z)}  dz^2  + 2 V(\rho,z) d\rho \, dz + e^{2 S(\rho,z)} d\Omega_{(2)}^2 
\ee
where the functions $r$, $U$, $V$, $S$ and $V$ depend only on $\rho$ and $z$ due to the axisymmetry, and these coordinates can be found such that the horizon is at the location $\rho= 0$, so that $r(0,z) = 0$, and that $z$ is an angle with the usual period of $2\pi$, so $z \sim z + 2 \pi$. 
From our second assumption we see that $S(\rho,z)$ should be bounded from below, so that the 2-sphere is always of finite size.
The cavity boundary is then defined by the condition that $r(\rho,z) = 1$ which implicitly selects a curve in the $\rho$ and $z$ coordinates.
We next assume that the  metric in this chart is a fibration over the 2-dimensional Riemannian base manifold defined by taking a point on the Euclidean time circle and on the 2-sphere,  so having the metric, $ds^2_{2d} = e^{2 U(\rho,z)} d\rho^2 + e^{2 W(\rho,z)}  dz^2  + 2 V(\rho,z) d\rho \, dz$, so that the horizon and cavity are both boundaries of this, and the determinant of this base metric is strictly positive.
Regularity of the Euclidean horizon at $\rho = 0$ determines all the functions  $U$, $W$, and $S$ to be smooth functions of $\rho^2$ at the horizon, and,
\be
r(\rho,z) = \rho \, \tilde{r}(\rho,z) \; , \quad V(\rho,z) = \rho \, \tilde{V}(\rho,z)
\ee
where $\tilde{r}$ and $\tilde{V}$ are again smooth functions of $\rho^2$~\cite{Headrick:2009pv,Wiseman:2011by}. It further requires that,
\be
\label{eq:beta}
\left. \frac{e^{U}}{\tilde{r}} \right|_{\rho = 0} = \frac{\beta}{2 \pi}
\ee
which is the Euclidean version of the constant surface gravity condition, and in particular relates the $z$ dependence of the functions $\tilde{r}$ and $U$ at the horizon.
Our final assumption is that there exists a reflection symmetry in the plane $z = 0$ so that,
\be
z \to - z \quad \implies \quad ( r, U, W , S, V ) \to ( r, U, W , S, - V ) \; .
\ee
All the known homogeneous and inhomogeneous black hole solutions (and indeed also the localized black holes) have such a reflection symmetry -- it may be that there are as yet undiscovered solutions that lack it. 

\subsection{The Harmark-Obers Ansatz}

This general metric contains 5 unknown functions. Using the freedom to perform a diffeomorphism, $(\rho,z) \to (r, \theta)$ with
$r = r(\rho,z)$ and $\theta = \theta(\rho,z)$ we may hope to reduce this to 3 unknowns at least locally. Let us assume for the moment that we may simplify the metric making the coordinate choice,
\be
\label{eq:startingansatz}
ds^2 =  r^2 T^2(r,\theta) d\tau^2 + e^{2 A(r,\theta)} \left( \Phi(r,\theta) dr^2 + L^2 d\theta^2 \right) + \sqrt{\Phi(r,\theta)} d\Omega_{(2)}^2  
\ee
where $\theta$ is an angular coordinate so $\theta \sim \theta + 2 \pi$.
Provided that $T(r,\theta)$ is a non-zero smooth function of $r^2$, then $r = 0$ is a black string horizon (assuming the other functions are also smooth in $r^2$ and the surface gravity condition is obeyed).
Here we have eliminated the off-diagonal component and imposed a relation between the diagonal spatial components. The remarkable property of this coordinate choice is that the time-time component of the Ricci flatness condition becomes simply,
\be
\label{eq:Teqn}
  \partial_r^2 T + \frac{2}{ r} \partial_r T + \frac{1}{L^2} \partial_\theta \left( \Phi \partial_\theta T  \right)  = 0 \; .
\ee
This may be solved by the choice $T(r,\theta) = $constant. Choosing the constant $T = 1$ and then taking the coordinate domain as $r \in [0, 1 ]$ precisely yields a horizon at $r = 0$ and a cavity wall at $r = 1$ where $g_{\tau\tau} = 1$ as required, and now the metric is simply given in terms of two unknown functions $\Phi(r,\theta)$ and $A(r,\theta)$ so that,
\be
\label{eq:ansatz}
ds^2 =  r^2 d\tau^2 + e^{2 A(r,\theta)} \left( \Phi(r,\theta) dr^2 + L^2 d\theta^2 \right) + \sqrt{\Phi(r,\theta)} d\Omega_{(2)}^2  \; .
\ee
Turning this around, if we say that our coordinate domain is $r \in [0, 1]$, with the cavity at $r = 1$, then an application of the maximum principle implies that  $T^2 = 1$ is the unique solution to~\eqref{eq:Teqn}.\footnote{
There appear to be two boundaries for this equation, one at $r = 1$ and the other at $r =0$. However this second horizon boundary can be removed by lifting the $r$ coordinate to be the radial coordinate of 3d Euclidean space, in a manner similar to that which we use later in section~\ref{sec:EOM}.
} 
Of course if the boundary condition at the cavity wall $r = 1$ was that $g_{\tau\tau}$ was not constant, then $T$ would in general be a complicated function of $r$ and $\theta$.

The key point of this form of the metric is that we may now measure the inhomogeneity of a solution naturally with respect to the lapse function simply by considering the derivative of its metric functions $\Phi$ and $A$ with respect to $\theta$, noting that this vanishes for the lapse function. In particular we may decompose the two remaining metric functions in terms of a homogeneous zero-mode and Kaluza-Klein modes as,
\be
{\Phi}(r, \theta) = \bar{\Phi}(r) + \sum_{n \ne 0} \Phi_n(r) e^{i n \theta}
\ee
and similarly for $A(r, \theta)$. We are then interested in the behaviour of the Kaluza-Klein modes as the circle size becomes small relative to some measure of the size of the black hole. 

So far in the discussion  we have assumed that we could consistently choose the ansatz~\eqref{eq:startingansatz}, and then showed that for our boundary conditions, the simpler form~\eqref{eq:ansatz} applies. How can we be sure that we may find this coordinate system for a solution, and further that it globally covers it?
This is particularly important if we wish to prove that inhomogeneous black strings cannot exist for small circle size -- if we were to prove that no inhomogeneous solution to the Einstein equations involving $\Phi$ and $A$ exists for small circle size without a proof that the ansatz~\eqref{eq:ansatz} can always be chosen, then we could not be sure that our statement of non-existence of inhomogeneous solutions applies to the black holes themselves, or simply to the existence of a global coordinate chart of the  particular form~\eqref{eq:ansatz}.
We now prove that we may indeed construct the diffeomorphism from our general metric~\eqref{eq:generalmetric}, which we have assumed to exist, to this special form.

\subsection{Coordinate transformation}

Since the metric~\eqref{eq:static} is a solution of the vacuum Einstein equations, the Ricci flatness condition implies that the lapse function $r(\rho,z)$ is harmonic with respect to the 4-dimensional spatial metric $g_{ij}$, so,
\be
\label{eq:rharmonic}
\nabla^2 r = \frac{1}{\sqrt{g}  } \partial_i \left( \sqrt{g} \, g^{ij} \partial_j r(x) \right) = 0 \; .
\ee
We proceed by writing the determinant of the base metric, the 2 by 2 block given by the $\rho$ and $z$ coordinates, as $\Delta = \sqrt{e^{2 (U + W)} - V^2}$, noting from our assumptions that this must be strictly positive, $\Delta > 0$.
We then consider a function $\Theta(\rho,z)$ defined as the solution of the system,
\be
\label{eq:theta}
\partial_\rho \Theta(\rho,z) = \frac{e^{2 S(\rho,z)}}{\Delta(\rho,z)} \left( e^{2 U(\rho,z)} \partial_z r(\rho,z) - V(\rho,z) \partial_\rho r(\rho,z) \right) \; , \quad 
\partial_z \Theta(\rho,z) = - \frac{e^{2 S(\rho,z)}}{\Delta(\rho,z)} \left( e^{2 W(\rho,z)} \partial_\rho r(\rho,z) - V(\rho,z) \partial_z r(\rho,z) \right) \; . \nl
\ee
The compatibility condition for this pair of differential equations is precisely the harmonic condition~\eqref{eq:rharmonic}. Hence provided we have a Ricci flat metric, we may construct the function $\Theta$ by solving these. We do so by picking the function $\Theta$ to vanish at the point $(\rho,z) = (0,0)$. Then we integrate the second condition along the horizon,
\be
\partial_z \Theta(0,z) = - \frac{e^{2 S(0,z)}}{\Delta(0,z)} e^{2 W(0,z)} \tilde{r}(0,z) 
\ee
 to construct $\Theta(0,z)$. We then find $\Theta(\rho,z)$ by integrating in $\rho$ at constant $z$ using the first condition above with $\Theta(0,z)$ as initial data. We note that since $\partial_z r = 0$ and $V =0$ at $z = 0$, then $\Theta(\rho,0) = 0$ and likewise $\Theta(\rho,2\pi)$ is constant; let us denote this constant as $\Theta(\rho,2\pi) = \Theta_0$.  
 
 We note that if the cavity boundary is a curve in the $\rho$-$z$ plane that bends back on itself, so that it intersects a constant $z$ line multiple times, then this simple algorithm will fail, but nonetheless $\Theta$ can still be solved by integration in the domain, although not along lines of constant $z$. We emphasize that the equations~\eqref{eq:theta} allow $\Theta$ to be integrated along any curve.
 
Having constructed $\Theta(\rho,z)$, we then use this together with $r(\rho,z)$ to construct the diffeomorphism,
\be
( \rho, z ) \to ( r = r(\rho,z), \Theta = \Theta(\rho,z) ) \; .
\ee
We may check that this diffeomorphism is locally invertible; the inverse function theorem states that in the neighbourhood of a point invertibility requires that the determinant,
\be
Q \equiv \det 
\left(
\begin{array}{cc}
\partial_\rho r & \partial_z r \\
\partial_\rho \Theta & \partial_z \Theta
\end{array}
\right) 
\ee
does not vanish. Using the differential conditions, we may write this as,
\be
Q = - \frac{e^{2 S}}{\Delta} \matvec{v} \cdot \mat{M} \cdot  \matvec{v} 
\ee
where,
\be
\label{eq:Mandv}
\mat{M} = \left(
\begin{array}{cc}
e^{2 U} & V  \\
V  & e^{2 W}
\end{array}
\right)
 \; , \quad  \matvec{v} = \left( - \partial_z r,  \partial_\rho r  \right) \; .
\ee
Now since $\mat{M}$ is required to be strictly positive for the metric to be Euclidean, then we see that if $| \matvec{v} | > 0$, so that the gradient of $r$ does not vanish, then
 $\matvec{v} \cdot \mat{M} \cdot  \matvec{v} > 0$, and hence $Q$ can not vanish.

 We now argue that indeed the gradient of $r$ cannot vanish. The function $r$ is periodic in $z$ and a harmonic function with respect to the 4-dimensional spatial geometry due to~\eqref{eq:rharmonic}. We may regard it as a function on the cylindrical surface formed by the angle $z$ and the coordinate $\rho$, with one end defined by $\rho = 0$ where $r = 0$, and the other end defined by the locus where $r = 1$. Now since $r$ is harmonic, it cannot have a maximum or minimum within the interior, so $r^{-1}([0,1])$ forms a cylinder.
We note that the end where $r = 1$ may be a complicated shape in the $\rho$ and $z$ coordinates, but nonetheless defines the end of the cylinder.
The Hopf maximum principle applied to~\eqref{eq:rharmonic} then states that any maxima must occur at the ends of the cylinder, and further that the outer normal gradient must be strictly positive at a maximum there unless the function is everywhere constant. An analogous statement holds for  minima, where now the outer normal gradient must be strictly negative unless the function is constant. Now $r$ is not a constant function on the cylinder, as it takes different values at the two ends of the cylinder. However at each end of the cylinder $r$ is constant.
 This implies that the outer normal gradient cannot vanish anywhere on the ends of the cylinder, where $r$ takes its minimum and maximum values. In particular at the $r = 0$ end (the horizon) it must be everywhere strictly negative, and strictly positive everywhere at the $r=1$ end. 
 
 Now we consider the level sets of $r$ on the cylinder. Noting that the gradient of $r$ cannot vanish at the ends of the cylinder, this implies that level sets of $r$ near the ends must be homologous to the ends, and thus be circles that wrap the cylinder. 
A stationary point of $r$ where its gradient vanishes is characterized by either being a maxima, minima or a saddle.
Now since $r$ is harmonic there are no maxima or minima. However a harmonic function may have saddle points, and in principle, these may be degenerate (meaning that the Hessian of the function vanishes at the saddle point).
Let us firstly give a simple argument that no saddle points may exist in our setting in the generic case that all putative saddle points are non-degenerate. This builds intuition for why no saddles exist, and hence no critical points, but it is not a general argument.
We then appeal to a more sophisticated theorem for elliptic differential operators in two dimensions for the general case where saddles are allowed to be degenerate.

Let us consider the case where all critical points, here saddles, are non-degenerate. Then a straightforward application of Morse theory relates the number of each type of stationary point to the topology of the space. For our cylindrical surface it states,
 \footnote{The function $r$ is defined on a space homeomorphic to $[0,1] \times S^1$ and satisfies $r(0,x)=0$ and $r(1,x)=1$ for all $x \in S^1$ under this identification.  
 Since $r = 0$ is the minimum value, and $r = 1$ is the maximum, and the gradient at these ends is non-vanishing,
 we can extend $r$ to a function on $ [-1/2, 3/2] \times S^1$ satisfying $r(-1/2,x)=r(3/2,x)$ for all $x \in S^1$, and introducing one new maximum, one new minimum, and two new saddle points. We can then view it as a function on the $2$-torus and apply the Poincaré–Hopf theorem to it.}
 \be
 \label{eq:morse}
\left( \mathrm{\#maxima} \right) + \left( \mathrm{\#minima} \right) - \left( \mathrm{\#saddles} \right) = 0 
 \ee
 where these are understood to correspond to (non-degenerate) stationary points -- so they do not include the maxima and minima at the ends of the cylinder, since the gradient does not vanish there.
 Thus since $r$ has no maxima or minima, we learn that $r$ can have no stationary points, and thus its gradient cannot vanish anywhere.
 
 Now let us give the full argument that allows for the possibility that saddle points are degenerate. The function $r$ is harmonic on the full 4-dimensional spatial geometry, but only depends on the two dimensions $\rho$ and $z$, and we may reduce it to the solution of an elliptic equation on the 2-dimensional base.
 This base is a cylinder, covered by the chart given by the coordinates $\rho$ and $z$. We now give a diffeomorphism that maps this cylinder to an annulus in the plane with Cartesian coordinates $x$ and $y$, taking,
\be
x = (1 + \rho) \cos(z) \; , \quad y = (1 + \rho) \sin(z) 
\ee 
so that the horizon where $r = 0$ and $\rho =0$ is the circle $x^2 + y^2 = 1$, and the cavity boundary where $r = 1$ is a closed curve, homologous to the horizon circle, such that $x^2 + y^2 > 1$.
Now the harmonic condition~\eqref{eq:rharmonic} can be translated to the elliptic equation on this annulus as,
 \be
 \label{eq:elliptic}
\left( G^{AB} \partial_A \partial_B + V^A \partial_A \right) r(x,y) = 0
 \ee
 where $x^A = \{ x, y \}$ and,
\be
 \label{eq:elliptic2}
 G^{AB} =  \frac{1}{( 1 + \rho) \Delta} \mat{R}^T \cdot  \left( \begin{array}{cc}
 e^{2 W} & - V \\ - V & e^{2 U}
 \end{array} \right) \cdot \mat{R} \quad , \qquad \mat{R} =  \left( \begin{array}{cc}
\cos(z) & \sin(z)  \\ - ( 1 + \rho ) \sin(z) & ( 1 + \rho ) \cos(z)
 \end{array} \right) 
 \ee
 so that $G^{AB}$ is a rescaling of the push-forward of the inverse metric of the 2-dimensional base manifold under the diffeomorphism above,
 and $V^A$ involves the various functions entering the spatial metric $g_{ij}$ and their derivatives. We note that $G^{AB}$ is smooth since $\Delta > 0$ everywhere, and $\rho \ge 0$ in the annulus, and further obeys $\det(G^{AB}) = 1$.
The level sets of the function $r$ in the annulus are then governed by the main theorem in~\cite{alessandrini1992index} which is reviewed in~\cite{magnanini2016introductionstudycriticalpoints}. Suppose a function obeys an elliptic differential equation of the form~\eqref{eq:elliptic} in a bounded domain of the plane, whose boundaries form $J$ simple (non-intersecting) closed curves, and where the function value on these boundaries are given by prescribed constants. The theorem then states that the function has a finite number of critical points $z_1, \ldots , z_K$, and these obey the relation,
\be
\sum_{z_k} m(z_k) = J - 2
\ee
provided that there are no critical points on the boundaries of the domain, where $m(z_k)$ is the multiplicity of the critical point at $z_k$, where the multiplicity is a positive integer that provides a measure of the degeneracy of the critical point. 
By definition, non-degenerate critical points have multiplicity one, and degenerate critical points have a multiplicity greater than one.
We note there is a modification of this if critical points occur on the boundaries but we shall not require that here.
We may now apply this theorem to our situation where we have the annulus, and hence $J = 2$, and the elliptic problem for the function $r$ given by equations~\eqref{eq:elliptic} and~\eqref{eq:elliptic2}, and we have the Dirichlet conditions that $r = 0$ and $1$ on the boundaries of the annulus. Further we know that there are no critical points on the ends of the annulus (as we have shown that the gradient does not vanish there). The theorem then dictates that there can be no critical points in the annulus, and hence there are none in our chart formed by $\rho$ and $z$.
 Thus we learn that indeed $Q$ can not vanish and thus the pair of functions $(r, \Theta)$ form a good coordinate transformation locally. 
 Since level sets of the smooth function $r$ can only intersect at critical points, we also learn that the level sets of $r$ cannot cross each other, and hence must form a foliation of the cylinder, a fact we will use shortly.
 
It is convenient to rescale the periodic coordinate $\Theta$ as,
\be
\Theta = L \theta
\ee
so that $\theta \sim \theta + 2 \pi$ is an angle, and then $L = \Theta_0/(2\pi)$. Now writing the metric using these $(r, \theta)$ coordinates we find our ansatz~\eqref{eq:ansatz}
where we have,
\be
\Phi(r,\theta) = \left. e^{S(\rho,z)} \right|_{\rho = \rho(r,L \theta),\,z = z(r,L \theta)}  \; , \quad
e^{2 A(r,\theta)} = \left. \frac{ e^{-4 S(\rho,z)} \Delta^2  }{  \matvec{v} \cdot \mat{M} \cdot  \matvec{v} } \right|_{\rho = \rho(r,L \theta),\,z = z(r,L \theta)} 
\ee
and the domain of the chart is $r \in [ 0, 1 ]$ with $r=1$ being the cavity boundary. We note that since $\Delta > 0$ and $\matvec{v} \cdot \mat{M} \cdot  \matvec{v}  > 0$, then $\Phi$ and $A$ are smooth functions in the domain.

Crucially we must check that the coordinate transform is globally well defined. 
In particular  the issue of global invertibility of the coordinates arises in the above when we construct the functions $\Phi(r, \theta)$ and $A(r, \theta)$.
Firstly we may see that the map $(\rho,z) \to (r, \theta)$ is well defined with the target domain being defined by $r \in [0 , 1]$ and $\theta$ as an angle. Since $r(\rho,z)$ has no minima or maxima within the interior of the domain of $\rho$ and $z$, then its minimum and maximum values are the ones on the boundaries, so zero and one respectively. Likewise from~\eqref{eq:theta} the function $\Theta$ and hence $\theta$ can have no minima or maxima except at boundaries and hence in the domain $z \in [0, 2\pi]$ the minimum and maximum values of $\theta$ are zero and $2 \pi$. 
Thus the map is indeed well defined with the target being $r \in [0 , 1]$ and $\theta$ having the usual angular period.
Furthermore, since $r$ and $\theta$ are smooth, and hence continuous functions of $\rho$ and $z$, the map is a clearly surjection. 

However, we must show the map is an injection. Again consider the cylinder surface formed by $\rho$ and the angle $z$, and consider the level sets of $r$ which we have argued are circles wrapping the cylinder that foliate it. Consider one such level set. A unit tangent to it is given by taking the vector $\matvec{v}$ defined above in~\eqref{eq:Mandv} and normalizing it  as, 
\be
\hat{\matvec{v}} = \frac{1}{| \matvec{v} |} ( - \partial_z r , \partial_\rho r )
\ee
where we note that $| \matvec{v} |^2 = (\partial_\rho r )^2 + (\partial_z r)^2 > 0$ since we have shown the gradient of $r$ cannot vanish. Now consider how $\theta$ evolves along this curve. Its gradient along the level set curve is,
\be
\partial_{\hat{\matvec{v}}} \theta = \frac{1}{L | \matvec{v} |} \left(  \partial_\rho r \partial_z \Theta -  \partial_z r \partial_r \Theta  \right) = - \frac{1}{L | \matvec{v} |} \frac{ e^{2 S} }{ \Delta  }   \matvec{v} \cdot \mat{M} \cdot  \matvec{v} =  \frac{Q}{L | \matvec{v} |} > 0
\ee
and since $Q > 0$ and $|\matvec{v}|> 0$ then we see that $\theta$ is a monotonic function along the level sets of $r$. 
Thus since each value of $r$ is associated to a single level set wrapping the cylinder, and $\theta$ is a monotonic function on this level set, then as we wrap around from $z = 0$ to $z = 2\pi$ we have that no two points on the cylinder can share the same values of $r$ and $\theta$. Hence the map is injective, completing the argument.

\subsection{Moduli space and homogeneous black strings}

Our interest is in the existence of inhomogeneous black string solutions. Hence we should fix the cavity geometry to be homogeneous, so that any inhomogeneity is intrinsic to the infilling black string solution, rather than imposed by boundary conditions. We have already chosen the constant potential condition that $g_{\tau\tau} = 1$ at the cavity boundary. However we will need more boundary conditions to specify the infilling geometry. With this in mind it is interesting to understand how physical data about the cavity are determined;
\begin{itemize}
\item Given that  $g_{\tau\tau} = 1$, the proper length of the time circle at the cavity is $\beta$. However we emphasize that $\beta$ must be determined by the infilling solution via~\eqref{eq:beta} to be $\beta =  \left. 4 \pi e^{A} \sqrt{\Phi} \right|_{r=0}$ to ensure the Euclidean horizon is smooth.
\item The value of $\Phi$ determines the proper radius $R_{sph}$ of the cavity 2-sphere as $R_{sph} =   \left. \Phi^{\frac{1}{4}} \right|_{r=1}$, and we must take this to be constant over the cavity boundary to ensure the boundary conditions are compatible with homogeneous infilling solutions.
\item The value of $L$ and the average value of $A$ at the cavity determine the proper length of the Kaluza-Klein circle to be $ \left. L \int d\theta e^A \right|_{r=1}$.
\end{itemize}
Since we are considering the vacuum Einstein equations, the above metric in equation~\eqref{eq:ansatz} has a further invariance that allows us to shift the value of the function $A(r,\theta)$ by a constant. For any static metric of the form~\eqref{eq:static} with spatial metric $g_{ij}(x^k)$ that solves the vacuum equations, we can rescale the spatial metric by a constant $\lambda$ to $\lambda^2 g_{ij}(x^k)$ and obtain another  solution. We achieve this for our ansatz by taking,
\be
 L \to \lambda^2 L \; , \quad
e^A \to \frac{1}{\lambda} e^A \; , \quad \Phi \to \lambda^4 \Phi
\ee
which precisely scales the spatial metric in this way, and preserves our boundary condition $g_{\tau\tau} = 1$. Using this invariance we can choose to fix, 
\be
\label{eq:Acondition}
2 \pi = \left. \int d\theta e^A \right|_{r=1}
\ee
so that the cavity circle radius is simply $L$ and its size is  given by $2 \pi L$. 

It is instructive to look at the family of homogeneous black string solutions in our ansatz~\eqref{eq:ansatz}. 
If we restrict the metric functions to be functions of $r$ only, so that $A = A(r)$, $\Phi = \Phi(r)$, then
we find that the vanishing of $R_{\theta\theta}$ implies, $A''(r) + r A'(r) = 0$, and the only solution consistent with horizon regularity and the cavity condition~\eqref{eq:Acondition}  is that it vanishes, $A(r) = 0$.
  From the perspective of the Kaluza-Klein reduction to gravity and a scalar, we may think of this as the scalar equation, and uniqueness of the scalar dictates only the trivial solution. 

The vanishing of $R_{\Omega^A\Omega^B}$, where $\Omega^A$ are coordinates on the 2-sphere, then gives an equation for $\Phi$, which we find to be,
\be
\Phi''(r) + \frac{1}{r} \Phi'(r) - \frac{\Phi'(r)^2}{\Phi(r)} - 4 \Phi(r)^{\frac{3}{2}} = 0 \; .
\ee
At the horizon this has a singular behaviour going as $\Phi(r) \sim  \frac{c}{r^2} ( 1 + O(r) )$ where $c$ is a constant of integration. However the regular solution goes as,
\be
\Phi(r) = \left(  \frac{2 \mu}{1 - \mu^2 r^2 } \right)^4
\ee
where $\mu$ is the other constant of integration. The 4-dimensional section of this solution is then simply the Schwarzschild solution with horizon radius $r_0 = 2 \mu$, and since $A = 0$, the 5-dimensional solution is simply a product of this Schwarzschild metric with a circle of radius $L$ -- the homogeneous black string family -- and $\mu$ is the mass per unit length. We may compute $\beta$ from the regularity condition, and the proper cavity radius $R_{sph}$, finding that,
\be
\beta = 8 \pi \mu^2  \; , \quad R_{sph} = \frac{2 \mu}{1 - \mu^2  } \; .
\ee
Then  we obtain, $\beta/R_{sph}  = 4 \pi \mu (1 - \mu^2)$, and varying the modulus $\mu$ we scan across the small and large homogeneous black strings in the cavity.

Having fixed the cavity boundary conditions that $g_{\tau\tau} = 1$ and the proper size of the Kaluza-Klein circle is $L$, then a further boundary condition is required to control the mass per unit length of the infilling homogeneous string. It is natural to take this to be a boundary condition on $\Phi$  -- such as its value at the cavity wall, or a condition on its normal derivative -- which will then determine one (or a discrete set) of these homogeneous strings, and in particular its value of $\mu$, the mass per unit length.

\subsection{Equations and boundary conditions}
\label{sec:EOM}

The non-trivial vacuum Einstein equations for our ansatz~\eqref{eq:ansatz} are given by the $rr$, $\theta\theta$, $r\theta$ and 2-sphere components. We emphasize that $R_{tt}$ is automatically satisfied by our ansatz. Vanishing $R_{\Omega^A\Omega^B}$ and $R_{\theta\theta}$ imply;
\be
\label{eq:E12}
0 = E_1 &\equiv& \partial_r^2 \Phi + \frac{1}{r} \partial_r \Phi + \frac{\Phi}{L^2}  \partial_\theta^2 \Phi  - \frac{(\partial_r \Phi)^2}{\Phi} - 4 \Phi^{\frac{3}{2}} e^{2 A}   \nl
0 = E_2 &\equiv& \partial_r^2 A + \frac{1}{r} \partial_r A + \frac{1}{L^2} \left( \Phi \partial_\theta^2 A + \partial_\theta^2 \Phi  - \frac{5}{8} \frac{ (\partial_\theta \Phi)^2}{\Phi}  \right) 
\ee
which have an elliptic character. The remaining two equations $R_{rr}$ and $R_{r\theta}$ imply,
\be
0 = H_1 &\equiv& - 2 \Phi \left( R_{rr} + E_2 \right) = \partial_r^2 \Phi - \left( \frac{1}{r} + 2 \partial_r A + \frac{5}{4} \frac{\partial_r \Phi }{\Phi} \right) \partial_r \Phi - \frac{4 \Phi}{r} \partial_r A  - \frac{\Phi}{L^2} \left( \partial_\theta^2 \Phi - \frac{5}{4} \frac{ ( \partial_\theta \Phi )^2 }{ \Phi} - 2  \partial_\theta A  \partial_\theta \Phi  \right) \nl
0 = H_2 &\equiv& - 2 \Phi R_{r\theta} =  \partial_r \partial_\theta \Phi  - \frac{1}{r} \partial_\theta \Phi - \frac{2}{r} \Phi \partial_\theta A - \frac{5}{4} \frac{\partial_r \Phi \partial_\theta \Phi}{\Phi}  - \partial_r A \partial_\theta \Phi  - \partial_r \Phi \partial_\theta A
\ee
which both have a hyperbolic character. Now suppose we solve both the elliptic equations so that $E_1 = 0$ and $E_2 = 0$, thinking of these as equations that determine $\Phi$ and $A$. The obvious question is then how we can simultaneously also satisfy the two remaining equations $H_{1,2}$. If we define,
\be
\mathcal{R}_r = r R_{rr} \; , \quad \mathcal{R}_\theta = r R_{r \theta}
\ee
the contracted Bianchi identities imply,
\be
\label{eq:Bianchi}
\frac{1}{r} \partial_r \left( r \mathcal{R}_r \right) + \frac{1}{L^2} \partial_\theta \left( \Phi \mathcal{R}_\theta \right) = 0 \; , \quad  \partial_r \mathcal{R}_\theta - \partial_\theta  \mathcal{R}_r  = 0
\ee
which imply that,
\be
\label{eq:Rtheta}
\partial_r^2 \mathcal{R}_\theta + \frac{1}{r} \partial_r \mathcal{R}_\theta + \frac{1}{L^2}  \partial_\theta^2 \left(  \Phi \mathcal{R}_\theta \right) = 0 \; .
\ee
We further note that at the horizon, due to the smoothness of $A$ and $\Phi$ in $r^2$ then,
\be
\label{eq:surfacegrav}
\left. \mathcal{R}_\theta \right|_{r=0} = \left. \partial_\theta A + \frac{1}{2 \Phi} \partial_\theta \Phi \right|_{r=0}
\ee
which is the constant surface gravity condition, whereas $\mathcal{R}_r$ automatically vanishes. We also note that $\mathcal{R}_\theta$ is a smooth function of $r^2$ at the horizon, given that $\Phi$ and $A$ are too.

Now let us consider boundary conditions for the system whereby we must fix data for $\Phi$ and $A$.
One might think that fixing the sphere size, and hence taking Dirichlet data for $\Phi$ would be a natural choice. 
Indeed it is, but here we will instead choose to take a different boundary condition that will allow us to make a statement about inhomogeneity. 
We will choose $\Phi$ to be constant, but we will not prescribe its value. Instead we will give the value of the integral of its normal derivative over the boundary; thus our boundary conditions for $\Phi$ are,
\be
\label{eq:bc1}
 \left. \int d\theta \partial_ r \Phi \right|_{r=1} = \xi \; , \quad \left. \partial_\theta \Phi \right|_{r=1}   = 0
\ee
so that $\xi$ is a constant. 
We may think of $\xi$ as a measure of how rapidly the 2-sphere is expanding at the cavity wall -- for an empty cavity filled with flat space this would determine the cavity radius.
We must remember that we have 4 equations to solve, and we are only imposing $E_{1,2}$ directly. The remaining two are solved by virtue of the contracted Bianchi identity provided that $A$ obeys appropriate conditions at the cavity wall. 
Suppose we arrange that $\mathcal{R}_\theta = 0$ at the cavity wall. We might then hope that equation~\eqref{eq:Rtheta}  implies that ${\mathcal{R}}_\theta$ must vanish everywhere in the interior of the domain. However naively this would also require that the surface gravity condition~\eqref{eq:surfacegrav} must also be imposed at the $r=0$ horizon boundary. However in fact this is not the case, and the surface gravity condition follows automatically if ${\mathcal{R}}_\theta$ vanishes on the cavity wall as we now argue.

The most straightforward way to see this is to lift the equation to an auxiliary 3-dimensional space that is a product of the disc formed by Euclidean time and $r$, together with the Kaluza-Klein circle direction. On to this space we introduce the flat Euclidean metric, to obtain,
\be
ds^2_{aux'} = r^2 d\tau^2 + dr^2 + L^2 d\theta^2 = dq^A dq^A + L^2 d\theta^2
\ee
so that $q^A = (r \cos{\tau}, r \sin{\tau})$ are the Cartesian coordinates corresponding to the polar coordinates $(r, \tau)$. Now we may lift the function $\mathcal{R}_\theta(r,\theta)$ to a circularly symmetric -- so $\tau$ independent -- function $\tilde{\mathcal{R}}_\theta(q^A, \theta)$ on this auxiliary space. Since $\mathcal{R}_\theta$ is a smooth function of $r^2$ at $r=0$ then this guarantees that it is  also a smooth function when lifted. The equation~\eqref{eq:Rtheta} lifts to,
\be
\partial_{q^A}^2 \tilde{\mathcal{R}}_\theta  + \frac{1}{L^2}  \partial_\theta^2 \left(  \tilde{\Phi}  \tilde{\mathcal{R}}_\theta \right) = 0 
\ee
so that circularly symmetric solutions of this descend to give $\mathcal{R}_\theta$ for our Einstein equations. The advantage is that $r=0$ is no longer a boundary in this lifted space. Thus we may apply the maximum principle to show that if $ \tilde{\mathcal{R}}_\theta$ is zero on the cavity wall, it must be zero everywhere in the interior. Reducing back to $r$ and $\theta$ this implies $\mathcal{R}_\theta$ must vanish everywhere, including at $r=0$, hence imposing constant surface gravity.
It is interesting that the two elliptic equations, $E_1$ and $E_2$, can also be lifted to the same auxiliary space,
\be
\label{eq:elliptic3d}
0 = \tilde{E}_1 &\equiv&  \partial^2_{q^A} \tilde{\Phi} + \frac{\tilde{\Phi}}{L^2}  \partial_\theta^2 \tilde{\Phi}  - \frac{(\partial_{q_A} \tilde{\Phi})^2}{\tilde{\Phi}^2} - 4 \tilde{\Phi}^{\frac{3}{2}} e^{2 A}  \nl
0 = \tilde{E}_2 &\equiv& \partial_{q^A}^2 \tilde{A} + \frac{1}{L^2} \left( \tilde{\Phi} \partial_\theta^2 \tilde{A} + \partial_\theta^2 \tilde{\Phi}  - \frac{5}{8} \frac{ (\partial_\theta \tilde{\Phi})^2}{\tilde{\Phi}}  \right) 
\ee
although we will not use this fact here.

For $\mathcal{R}_\theta = 0$ then the system~\eqref{eq:Bianchi} implies that $\partial_r( r \mathcal{R}_r ) = 0$, and since $\mathcal{R}_r$ vanishes at the horizon, it must then vanish everywhere.
 Thus we choose our boundary condition for $A$ at the cavity wall to be the vanishing of $\mathcal{R}_\theta$, and this is sufficient to impose that  $\mathcal{R}_{r, \theta}$ are zero everywhere. 
  With our choice of boundary conditions above for $\Phi$,
 we find that this implies,
 \be
 \label{eq:bc2}
\left. \frac{ e^{+A} }{ 2 \Phi +  \partial_r \Phi } \right|_{r=1} = \mathrm{constant} \; .
 \ee
 We  fix the constant by requiring our cavity condition on $A$ given earlier in~\eqref{eq:Acondition}.
For the homogeneous black string with these boundary conditions on $A$ and $\Phi$ the constant $\xi$ takes the value,
\be
\xi_{homog} =  \frac{128 \mu^6  }{\left( 1 - \mu^2  \right)^5} 
\ee
and so fixing $\xi$ fixes the mass per unit length, $\mu$, of the black string.

\subsection{Constraint on Kaluza-Klein modes}

We are now in a position to deduce a bound on how inhomogeneous black string solutions may be, given the circle size $L$ and the data $\xi$ that we have prescribed, 
together with our constant potential condition, and requirement that the sphere is homogeneous.
Consider $E_1$ in equation~\eqref{eq:E12} that governs ${\Phi}$. Now noting that the last two terms of the righthand side are negative, then rearranging and integrating over the chart with the weight $r$ we obtain the inequality,
\be
 -  \frac{1}{L^2}  \int dr d\theta \, r  {\Phi} \partial_\theta^2 {\Phi}  \le  \int dr d\theta \, \partial_r \left( r  \partial_{r} {\Phi} \right) 
\ee
and then integrating by parts we find,
\be
  \int dr d\theta \, r ( \partial_\theta {\Phi} )^2  \le L^2 \left. \int  d\theta \partial_r {\Phi} \right|_{r = 1} =  2\pi L^2  \xi    \; .
\ee
Now let decompose ${\Phi}$ in Kaluza-Klein modes on the circle,
\be
{\Phi}(r, \theta) = \bar{\Phi}(r) + \sum_{n \ne 0} \Phi_n(r) e^{i n \theta}
\ee
then the above inequality becomes the following bound,
\be
4 \pi \sum_{n > 0}  n^2 \int dr\, r \left| \Phi_n(r) \right|^2 \le  2\pi L^2  \xi \quad \implies \quad \int dr \, r  \left| \Phi_n(r) \right|^2 \le \frac{L^2  \xi  }{ 2 n^2} \quad \forall \quad n \ne 0\; .
\ee
Thus we see that as we take $L \to 0$ fixing our data $\xi$, these norms of each Kaluza-Klein mode are forced to vanish; and in this sense the solution must become increasingly homogeneous.
We note that rather than being the $\mathcal{L}_2$ norm of $\Phi_n$, we find a weighted norm. If we repeat this analysis with $\tilde{\Phi}$ in the 3-dimensional auxiliary space, then we find the same bound but with the conventional $\mathcal{L}_2$ norm in this higher dimensional space. Thus we see the weight factor of $r$ in the bound above expresses that it more naturally lives on the lifted space.

Finally we see our main result. 
Taking our boundary conditions~\eqref{eq:Acondition},~\eqref{eq:bc1} and~\eqref{eq:bc2} together with the constant potential condition $g_{\tau\tau} = 1$, then fixing the data $\xi$,
which for homogeneous strings is equivalent to fixing the mass per unit length, then the weighted norm of the Kaluza-Klein modes goes to zero as we take the circle radius $L$ to zero, implying that black string solutions must become more homogeneous.

\section{Discussion}
\label{eq:discussion}

We have set out to investigate our proposed conjecture~\ref{conj} in the introduction. We introduced a toy scalar field model that lives on the Kaluza-Klein vacuum spacetime, and whose static solutions have a similar structure to those in the gravitational context. For this toy model, we may easily prove the analog to this conjecture once the theory is formulated in a cavity whose radius is held fixed as we consider taking a small circle size. However we also find a weaker result, a bound on the norm of Kaluza-Klein modes in terms of the circle size. It is this latter weaker statement that we are able to also formulate in the gravitational context.

As for the scalar field we place the Kaluza-Klein theory in a cavity where we specify the cavity boundary is homogeneous and fix the circle size and certain boundary data that, in the case of homogeneous black string solutions, fixes the mass per unit length. 
This corresponds to fixing a constant gravitational potential at the cavity wall, the size of the Kaluza-Klein circle $L$ together with fixing a quantity analogous (but not identical to) the expansion of the spatial round 2-sphere.
 It would be interesting to understand in future work whether the cavity is necessary or whether a similar result could be found in the asymptotically Kaluza-Klein case, where the actual mass per unit length is fixed. With the cavity it would also be interesting to understand whether a more physical boundary condition, such as fixing the cavity radius, or an integral of the energy density as defined from the Brown-York tensor, could instead be taken. 
 
We have focussed on the static black hole solutions that are axisymmetric, so spherically symmetric in the 4 extended directions.
An important subtlety is that for the gravitational Kaluza-Klein theory a notion of homogeneity is harder to define as one may always perform a coordinate transformation that makes a homogeneous solution appear inhomogeneous. Here we have employed a specific coordinate chart, given in equation~\eqref{eq:ansatz}, similar to that introduced by Harmark and Obers~\cite{Harmark:2002tr}. This metric ansatz uses the time-time component of the metric as a coordinate. One virtue of this is that it reduces the number of metric functions to only two, while usually with this symmetry we would naively expect three. Another virtue is that the circle coordinate now may be used to meaningfully measure inhomogeneity relative to  that of the lapse function, which by construction is homogeneous in this coordinate chart.
Assuming some global chart exists for a static axisymmetric black string solution, we have proved that a diffeomorphism to our ansatz may always be found. Then in this chart, we prove that the inhomogeneity of the metric component controlling the 2-sphere radius is controlled by the circle size. Specifically decomposing this metric component in harmonics on the circle, a weighted $L^2$ norm of each coefficient wavefunction (except for the zero mode) is bounded from above above by the circle size. 
Our proof that this special coordinate chart in equation~\eqref{eq:ansatz} can always be chosen is an important result; without it a result bounding inhomogeneity in that chart might simply imply a limitation on when that coordinate chart exists, and say nothing about the inhomogeneity of actual black string solutions.

An important limitation of this approach is that we are not able to address the localized solutions, only the black holes with string topology so that their horizon wraps the compact circle. This is due to the localized solution metric components being singular in our chart. It would be very interesting to understand whether a more sophisticated chart might be introduced to also allow a similar result for the localized case.

We have shown that the Kaluza-Klein modes of the sphere metric function, as defined relative to the lapse, are bounded by the circle size. Thus for small circle size the sphere becomes increasingly homogeneous as measured in the weighted $L^2$ sense of the Kaluza-Klein wavefunctions. However a function that is small in an $L^2$ norm does not necessarily have to be small pointwise, and it would be very interesting to see whether a stronger pointwise statement could be made. 

Finally, obviously what we seek is not a restriction on the magnitude of inhomogeneity, but rather a strict ruling out of inhomogeneity given a fixed mass per unit length and then sufficiently small circle size. Even restricted to string topology black holes this would be a very interesting result, and we hope that future work may make progress in this direction. 
In particular we have shown that the problem may be phrased as a pair of scalar functions $\tilde{\Phi}$ and $\tilde{A}$ obeying the coupled elliptic equations~\eqref{eq:elliptic3d}
on the 3-dimensional space formed from the product of a 2-ball with the Kaluza-Klein circle, and governed locally by the Euclidean metric. Provided boundary conditions for $\tilde{\Phi}$ and $\tilde{A}$ are taken so that $\tilde{\mathcal{R}}_\theta$ vanishes, then solutions which are spherically symmetric in the 2-ball descend to smooth black string solutions in our gravitational problem. 
The virtue of this  phrasing of the problem is that there is now no mention of the horizon, and the underlying geometry is simply the flat 3-dimensional cylinder formed by the 2-ball and the Kaluza-Klein circle.
Thus if we can prove  in this 3-dimensional system of two coupled scalar fields that there are no inhomogeneous solutions for sufficiently small circle sizes, we will have proven the same statement for our gravitational system.

Of course we might wonder whether the conjecture could be false. We believe this is unlikely, but it would be extremely interesting as it would indicate some new very exotic black holes. It would be particularly interesting if these turned out to be dynamically stable in which case it would call into question the fundamental assumption of effective field theory applied to extra dimensions, namely that 4-dimensional physics is recovered as the circle size becomes small.

\subsection*{Acknowledgments}
We are very grateful to James Lucietti for initial collaboration and valuable discussions.
This work is supported by STFC Consolidated Grant ST/T000791/1. EA is funded by STFC studentships.

\clearpage

\addcontentsline{toc}{section}{Bibliography}
\bibliography{refs}

\end{document}